\documentclass[12pt, a4paper]{article}
\pdfoutput=1
\usepackage{amsmath,amssymb,bm,epsfig,afterpage}
\usepackage{cite}
\usepackage{here}
\usepackage{color}
\usepackage{colortbl}
\usepackage{tabularx}
\usepackage{subcaption}

\usepackage{graphicx}
\usepackage{listings}
\usepackage{longtable}
\usepackage[utf8]{inputenc}

\renewcommand{\thefootnote}{\fnsymbol{footnote}}

\newcommand{\ka}{\kappa}

\newcommand{\ol}[1]{\overline{#1}}

\newcommand{\abs}[1]{\left|#1\right|}
\newcommand{\order}[1]{\mathcal{O}\left(#1\right) }

\newcommand{\la}{\lambda}

\newcommand{\Lcal}{\mathcal{L}}
\newcommand{\Mcal}{\mathcal{M}}

\newcounter{num}

\newcommand{\GeV}{\mathrm{GeV}}

\newcommand{\SM}{\mathrm{SM}} 
\newcommand{\CDF}{\mathrm{CDF}}
\newcommand{\PDG}{\mathrm{PDG}}
\newcommand{\met}{E_T^\mathrm{miss}}
\newcommand{\dam}{\Delta a_\mu}

\usepackage[colorlinks=true, linkcolor=black, citecolor=black,
urlcolor=black]{hyperref}

\begin{document}

\begin{titlepage}

\begin{flushright}
{\tt CTPU-PTC-22-05}
\end{flushright}
\vspace{1.2cm}

\begin{center}

{\Large
{\bf 
$W$ boson mass and muon $g-2$ \\ in a lepton portal dark matter model
}
}

\vskip 2cm

Junichiro Kawamura$^{1}$\footnote{jkawa@ibs.re.kr}, 
Shohei Okawa$^{2}$\footnote{okawa@icc.ub.edu}
and
Yuji Omura$^{3}$\footnote{yomura@phys.kindai.ac.jp}

\vskip 0.5cm

{\it $^1$ Center for Theoretical Physics of the Universe, Institute for Basic Science (IBS), Daejeon, 34051, Korea}\\[3pt]

{\it $^2$ 
Departament de F\'isica Qu\`antica i Astrof\'isica, Institut de Ci\`encies del Cosmos (ICCUB),
Universitat de Barcelona, Mart\'i i Franqu\`es 1, E-08028 Barcelona, Spain
}\\[3pt]

{\it $^3$
Department of Physics, Kindai University, Higashi-Osaka, Osaka 577-8502, Japan}\\[3pt]

\vskip 1.5cm

\begin{abstract}
We study a lepton portal dark matter model,
motivated by the deviation of the $W$ boson mass reported by the CDF collaboration.
We introduce vector-like leptons and a scalar dark matter (DM) which exclusively couples to the extra leptons and muon.
The one-loop corrections induced by the new particles can shift the $W$ boson mass. Besides, the discrepancy in the muon anomalous magnetic moment and
the DM density can simultaneously be explained by this setup,
if the vector-like lepton is lighter than 200 GeV 
and nearly degenerate with the DM particle. 
We also see that the constraints on such a light extra lepton from the collider experiments can be evaded due to the existence of the DM particle. 
\end{abstract}

\end{center}
\end{titlepage}

\renewcommand{\thefootnote}{\arabic{footnote}}
\setcounter{footnote}{0}

\section{Introduction}

The precision measurement of the electroweak (EW) interaction plays an important role in tests of new physics.
In the Standard Model (SM), a Higgs doublet field develops
a non-vanishing vacuum expectation value (VEV),
and then the $Z$ and $W$ gauge bosons acquire their masses from the VEV. 
The mechanism can be tested by
the prediction of the EW precision observables (EWPOs), 
such as the $\rho$ parameter, defined as $\rho := m_W^2/(m_Z \cos\theta_W)^2$, 
where $\theta_W$ is the weak mixing angle.    
It is well known that new fields which contribute to the EW symmetry breaking 
can, in general, affect to the $\rho$ parameter at the tree-level, 
and hence it can constrain new physics models. 
Besides, there are loop corrections to the EWPOs 
where new particles charged under the EW symmetry run in the loops.

The CDF collaboration recently announced a new result of the $W$ boson mass measurement, 
$m_W^\CDF = 80.4335~(94)~\GeV$~\cite{CDF:2022hxs}.  
This value disagrees with the combinations of the other previous measurements, 
$m_W^\PDG = 80.379~(12)~\GeV$~\cite{ParticleDataGroup:2020ssz} 
and the significance is $7\sigma$.  
This new experimental result may suggest that the $W$ boson mass deviates 
from the SM prediction 
due to the existence of new physics beyond the SM. 
It has already been pointed out 
that the deviation can be interpreted by the corrections 
to the oblique parameters~\cite{Lu:2022bgw,Strumia:2022qkt,Balkin:2022glu}. 
In particular, the $T$ parameter is relevant to this deviation. 
In order to explain this discrepancy in new physics models, 
there should be a new particle at the EW scale 
that is not neutral under the EW symmetry, 
and/or mixes with SM particles~\cite{Strumia:2022qkt,deBlas:2022hdk,Yang:2022gvz,Yuan:2022cpw,Athron:2022qpo,Lu:2022bgw,Fan:2022dck,Babu:2022pdn,Heckman:2022the,Gu:2022htv,Athron:2022isz,DiLuzio:2022xns,Asadi:2022xiy,Bahl:2022xzi,Paul:2022dds,Bagnaschi:2022whn,Cheng:2022jyi,Lee:2022nqz,Liu:2022jdq,Fan:2022yly,Sakurai:2022hwh,Balkin:2022glu,Biekotter:2022abc,Endo:2022kiw,Crivellin:2022fdf,Heo:2022dey,Han:2022juu,Ahn:2022xeq, 
Song:2022xts,Blennow:2022yfm,Cacciapaglia:2022xih,Tang:2022pxh,Zhu:2022tpr,Zheng:2022irz,Krasnikov:2022xsi,Arias-Aragon:2022ats,Du:2022brr,Du:2022pbp}. 
Such a light particle, however, tends to be excluded by the constraints from the LHC experiments if it decays to detectable SM particles. 
A way to avoid the limits is that an EW particle responsible for the $m_W$ anomaly 
decays to an undetectable particle so that the signals are effectively invisible.

In this work, 
we discuss the $W$ boson mass in an extended SM where two vector-like leptons and a real scalar dark matter (DM) are introduced.  
In this model,
the scalar DM couples to the SM leptons via Yukawa couplings 
involving vector-like leptons.  
In Ref.~\cite{Kawamura:2020qxo}, 
the authors show that the relic density of DM and another recent anomaly in muon anomalous magnetic moment, 
$\Delta a_\mu = 2.51~(50)\times 10^{-9}$~\cite{Bennett:2004pv,Aoyama:2020ynm,Abi:2021gix,Aoyama:2012wk,Aoyama:2019ryr,Czarnecki:2002nt,Gnendiger:2013pva,Davier:2017zfy,Keshavarzi:2018mgv,Colangelo:2018mtw,Hoferichter:2019gzf,Davier:2019can,Keshavarzi:2019abf,Kurz:2014wya,Melnikov:2003xd,Masjuan:2017tvw,Colangelo:2017fiz,Hoferichter:2018kwz,Gerardin:2019vio,Bijnens:2019ghy,Colangelo:2019uex,Blum:2019ugy}, can be explained simultaneously, when the DM specifically couples to muon.   
We shall show that the discrepancy in the $W$ boson mass can be solved in this model as well.

The rest of this paper is organized as follows. 
The $W$ boson mass in the lepton portal DM model is discussed in Sec.~\ref{sec-Wmodel}, 
and then the DM physics and its relation to $\Delta a_\mu$ are studied in Sec.~\ref{sec-DMM}. 
Section~\ref{sec-summary} is devoted to summarize this paper. 
Loop functions used in our analysis are shown in Appendix~\ref{sec-lfun}.

\section{$W$ boson mass in lepton portal DM model} 
\label{sec-Wmodel}

\subsection{Model} 
\label{sec-model} 

We briefly review our model in this section. 
The more details are shown in Ref.~\cite{Kawamura:2020qxo}. 
The terms relevant to new particles in the lagrangian are given by 
\begin{align}
 -\Lcal_\mathrm{NP} 
=&\ \frac{1}{2} m_X^2 X^2 + m_L\ol{L}_L L_R + m_E \ol{E}_L E_R  \\ \notag 
 &\ + \la_L \ol{\ell}_L X L_R + \la_R \ol{E}_L X \mu_R    
        + \ka \ol{L}_L H E_R + \ol{\ka} \ol{E}_L i\sigma_2 H^\dag L_R + h.c.,  
\end{align}
where $L_{L,R}$ and $E_{L,R}$ are $SU(2)_L$ doublet and singlet vector-like leptons, 
respectively. 
A real scalar DM field is denoted by $X$  which is a singlet under the SM gauge group. 
$\ell_L = (\nu_\mu, \mu_L)$, $\mu_R$ and $H$ 
are the SM leptons in the second generation and the Higgs doublet, respectively. 
We neglect couplings of the scalar DM to the SM Higgs boson in the scalar potential. 
In our setup, 
we assume that the vector-like leptons couple exclusively to the second generation leptons, 
so that the muon anomalous magnetic moment is explained without lepton flavor violations. We note that a $Z_2$ symmetry is assigned and only the vector-like leptons and the DM particle $X$ are odd to ensure the stability 
of the DM.

After the EW symmetry breaking, the vector-like lepton masses are given by 
\begin{align}
 \Mcal_E = 
\begin{pmatrix}
 m_L & \ol{\ka} v_H \\ \ka v_H & m_E  
\end{pmatrix}, 
\quad 
 \Mcal_N = m_L,   
\end{align}
where $v_H = 174~\GeV$ is the Higgs VEV.
Note that there is no mixing between the vector-like leptons and the SM leptons. 
We define the mass eigenstates and diagonalizing matrices as 
\begin{align}
\begin{pmatrix}
  E_L^\prime  \\ E_L 
\end{pmatrix}
= 
U_L 
\begin{pmatrix}
  E_{L_1} \\ E_{L_2}
\end{pmatrix}, 
\quad 
\begin{pmatrix}
  E_R^\prime \\ E_R 
\end{pmatrix}
= 
U_R 
\begin{pmatrix}
  E_{R_1} \\ E_{R_2}
\end{pmatrix}, 
\quad 
U_L^\dag \Mcal_E U_R = \mathrm{diag}\left(m_{E_1}, m_{E_2}\right). 
\end{align}
and we parametrize the diagonalizing matrices as 
\begin{align}
 U_L = 
\begin{pmatrix}
 c_L & s_L \\ -s_L & c_L
\end{pmatrix},
\quad 
 U_R = 
\begin{pmatrix}
 c_R & s_R \\ -s_R & c_R
\end{pmatrix},
\end{align}
where $c_A^2 + s_A^2 = 1~(A=L,R)$. 
$E_{L,R}^\prime$ is a charged component in the doublet $L_{L,R}$. 
The neutral component in the gauge basis, $N_{L,R}$, 
is already in the mass basis $N_1$.

\subsection{Muon anomalous magnetic moment}

In this model, the new physics contribution to the muon anomalous magnetic moment, 
$\Delta a_\mu$, 
is originated from the 1-loop effects induced by the Yukawa couplings of muon with the vector-like leptons and $X$.  
$\Delta a_\mu$ is evaluated as ~\cite{Dermisek:2013gta,Jegerlehner:2009ry,Lynch:2001zs} 
\begin{align}
\label{eq-Delamu}
 \Delta a_\mu =  \frac{m_\mu}{16\pi^2 m_X^2} 
                  \Bigl[&  \left( c_R^2 |\la_L|^2 + s^2_L |\la_R|^2 \right) m_\mu F(x_1)
             + c_R s_L \text{Re} \left( \la_L \la_R \right) m_{E_1} G(x_1)    \\ \notag 
           &\quad              
 +  \left( s_R^2 |\la_L|^2 + c^2_L |\la_R|^2 \right) m_\mu F(x_2)
    - c_L s_R \text{Re} \left( \la_L \la_R \right) m_{E_2} G(x_2)\Bigr], 
\end{align}
where $x_i = m_{E_i}^2/m_X^2~(i=1,2)$. 
The functions $F$ and $G$ are defined in Appendix~\ref{sec-lfun}. 
For $m_{E_i} = \order{100}~\GeV$,  
$s_L\ne 0$ or $s_R \ne 0$ is required to explain the discrepancy 
due to the chirality enhanced effect. 
This means that both singlet and doublet vector-like leptons are necessary 
to explain $\Delta a_\mu$.

\subsection{Oblique parameters}
\label{sec-oblique}

New physics effects to the EWPOs are well described by the oblique parameters~\cite{Peskin:1990zt,Peskin:1991sw}. 
The one-loop contribution of the vector-like leptons, $E_1$, $E_2$ and $N_1$, to the oblique parameter $T$ 
is given by~\cite{Lavoura:1992np},  
\begin{align}
 16\pi s_W^2 c_W^2 T 
=&\ \left(c_L^2+c_R^2\right)\theta_+(x,y_1) 
  + 2 c_L c_R \theta_- (x,y_1)  \\  \notag 
   &\ + \left(s_L^2+s_R^2\right)\theta_+(x,y_2) 
 + 2 s_L s_R \theta_- (x,y_2)  \\ \notag 
   &\ -\left(c_L^2s_L^2 + c_R^2s_R^2 \right) \theta_+(y_1,y_2) 
       - 2c_Ls_L c_R s_R \theta_- (y_1,y_2), 
\end{align}   
where $x = m_{N_1}^2/m_Z^2$ and $y_i = m_{E_i}^2/m_Z^2$. 
Here, $c_W = \cos\theta_W$ and $s_W = \sin\theta_W$. 
The formula for $-2\pi U$ is given 
by replacing the loop functions $\theta_\pm \to \chi_\pm$. 
The one-loop contribution to the $S$ parameter is given by   
\begin{align}
 2\pi S 
=&\ \left(c_L^2+c_R^2\right)\psi_+(x,y_1) 
  + 2 c_L c_R \psi_- (x,y_1)  \\  \notag 
   &\ + \left(s_L^2+s_R^2\right)\psi_+(x,y_2) 
 + 2 s_L s_R \psi_- (x,y_2)  \\ \notag 
   &\ -\left(c_L^2s_L^2 + c_R^2s_R^2 \right) \chi_+(y_1,y_2) 
       - 2c_Ls_L c_R s_R \chi_- (y_1,y_2).  
\end{align}   
The loop functions $\theta_\pm$, $\chi_\pm$ and $\psi_\pm$ 
are defined in Appendix~\ref{sec-lfun}.

\begin{figure}[th]
\centering 
\begin{minipage}[c]{0.48\hsize}
 \centering
\includegraphics[height=65mm]{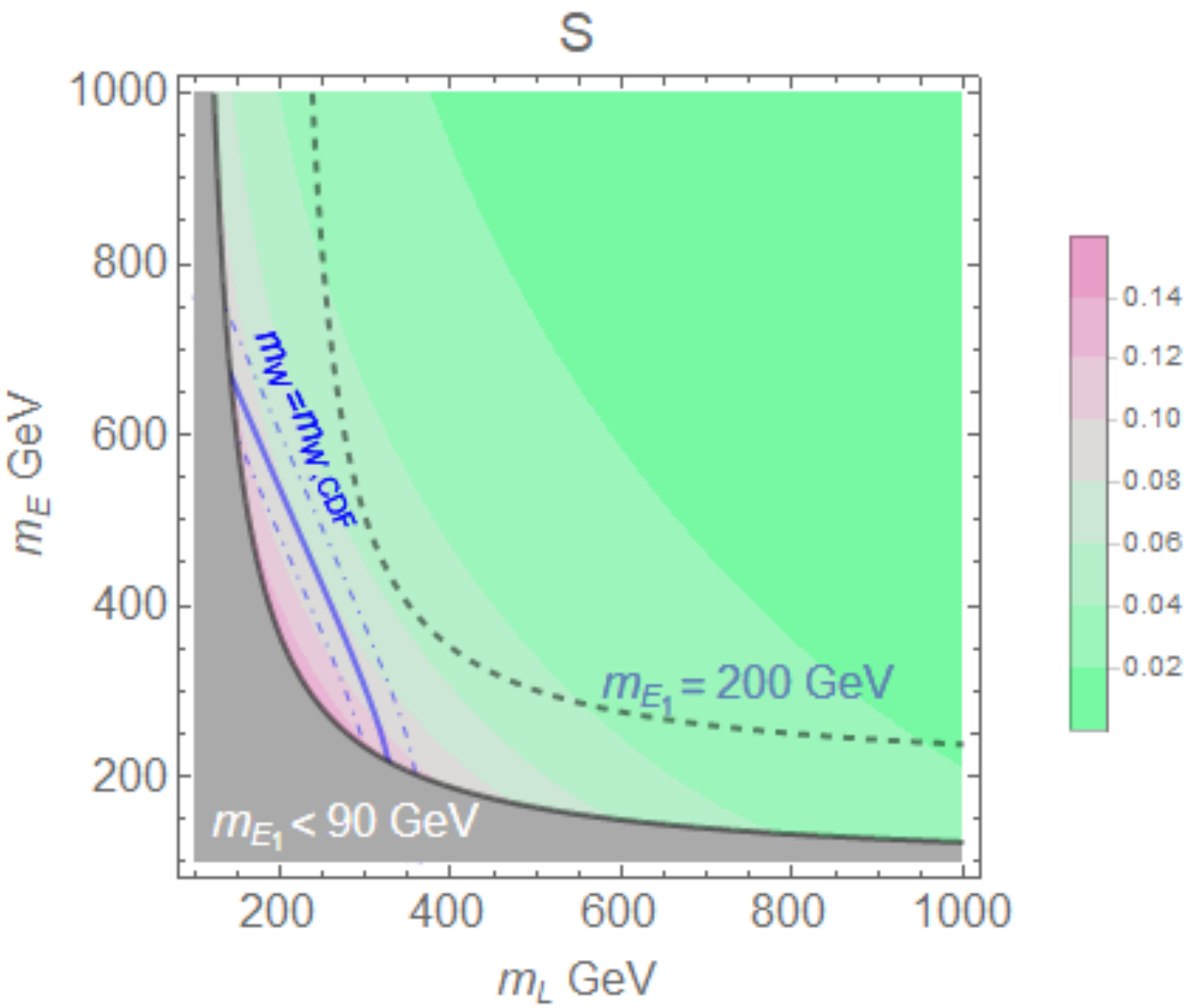} 
\end{minipage}
\begin{minipage}[c]{0.48\hsize}
 \centering
\includegraphics[height=65mm]{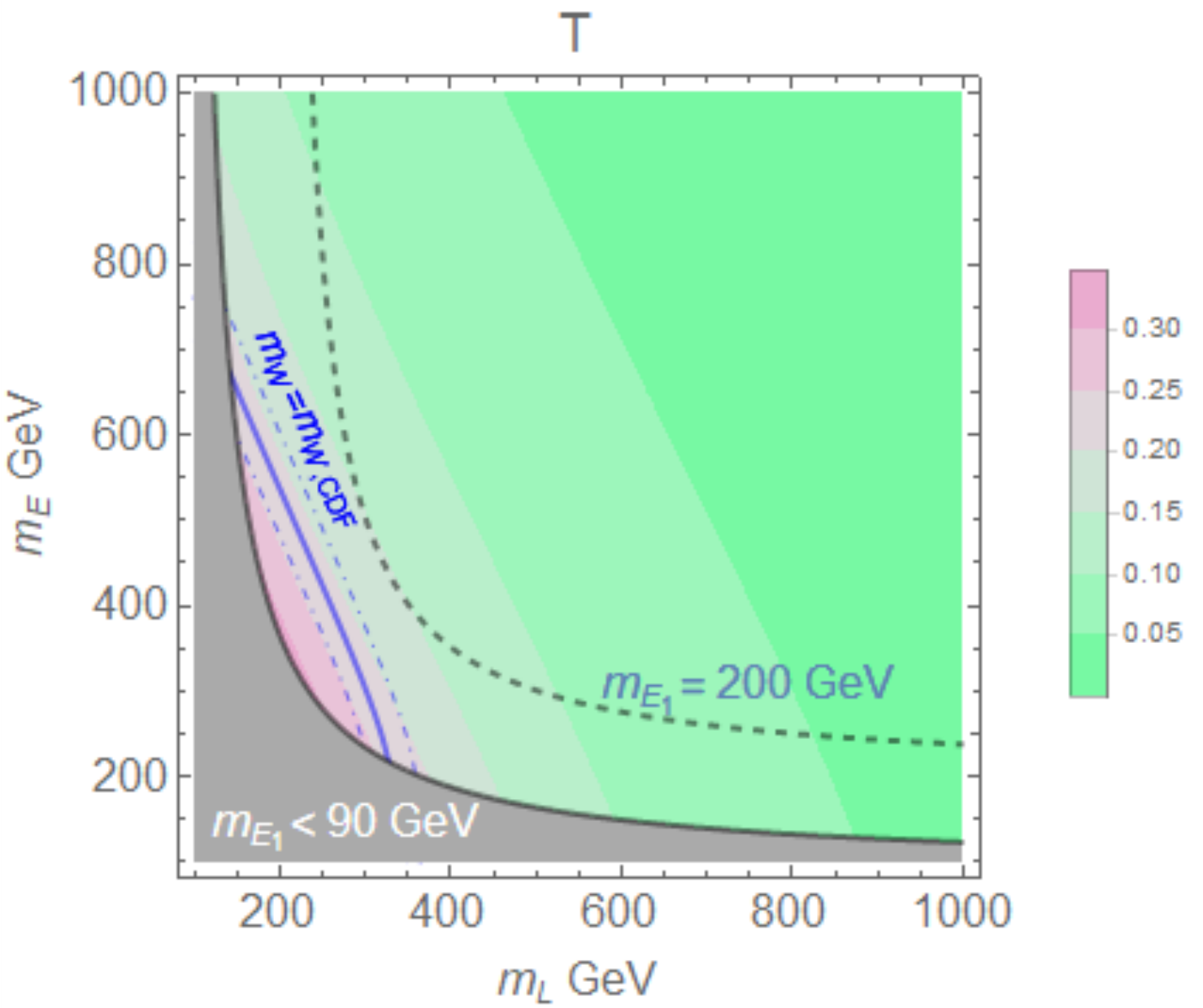} 
\end{minipage}
\caption{\label{fig-ST}
Values of $S$ and $T$ parameters for $\ka=\ol{\ka}=1.0$ 
with varying the vector-like masses. 
}
\end{figure}

Figure~\ref{fig-ST} shows the $S$ and $T$ parameters on the left and right 
panels, respectively. The horizontal and vertical axes correspond to
$m_L$ and $m_E$, respectively.
We take $\ka=\ol{\ka}=1.0$ in this figure. 
The $W$ boson mass reported by the CDF collaboration 
is explained on the blue line.
Note that the deviation of the $W$ boson mass from the SM prediction 
is related to the oblique parameters as~\cite{Maksymyk:1993zm,Grimus:2008nb} 
\begin{align}
  \frac{\delta m_W^2}{m_W^2|_\SM} =&\ \frac{\alpha}{c_W^2-s_W^2} 
    \left(  -\frac{1}{2} S + c_W^2 T + \frac{c_W^2-s_W^2}{4s_W^2} U\right) 
    \sim -0.007 S + 0.011 T + 0.0087 U, 
\end{align}
where $\alpha = 1/128$ and $s_W^2 = 0.22337$ are used in the second equality. 
The dot-dashed lines correspond to $1\sigma$ range of the CDF result. 
The lightest vector-like lepton mass is less than 90 GeV in the gray region, 
and is 200 GeV on the dashed line. 
As discussed later, the vector-like lepton lighter than 90 GeV 
may be excluded by the LEP experiment. 
$T \gtrsim 0.15$ can be realized by the light vector-like lepton, $m_{E_1} \lesssim 200~\GeV$. $S$ is positive in our parameter region.  
We also find $U < 0.03$ which is much smaller than the other oblique parameters.  

We also compare our predictions of the oblique parameters with 
the results based on the CDF measurement and the previous works in the PDG. 
In Ref.~\cite{Lu:2022bgw}, 
the EW-fit with the new CDF measurement and the PDG value are studied. 
The favored values of the oblique parameters with $U=0$ are given by 
$(S,T) = (0.15\pm 0.08, 0.27\pm 0.06)$ with the correlation coefficient $0.93$ 
for the analysis with the CDF result 
and 
$(S,T) = (0.05\pm 0.08, 0.09\pm 0.07)$ with the correlation coefficient $0.92$ 
for the analysis based on the PDG data. 
Similar values are obtained in Refs.~\cite{Strumia:2022qkt,Balkin:2022glu}. 
In the SM, $S=T=U=0$, 
the values of $\chi^2$ 
are 60 and 3.7 with the CDF and PDG data, respectively. 
Thus, the new data strongly favors the existence of new physics.

\begin{figure}[t!]
\centering 
\includegraphics[height=80mm]{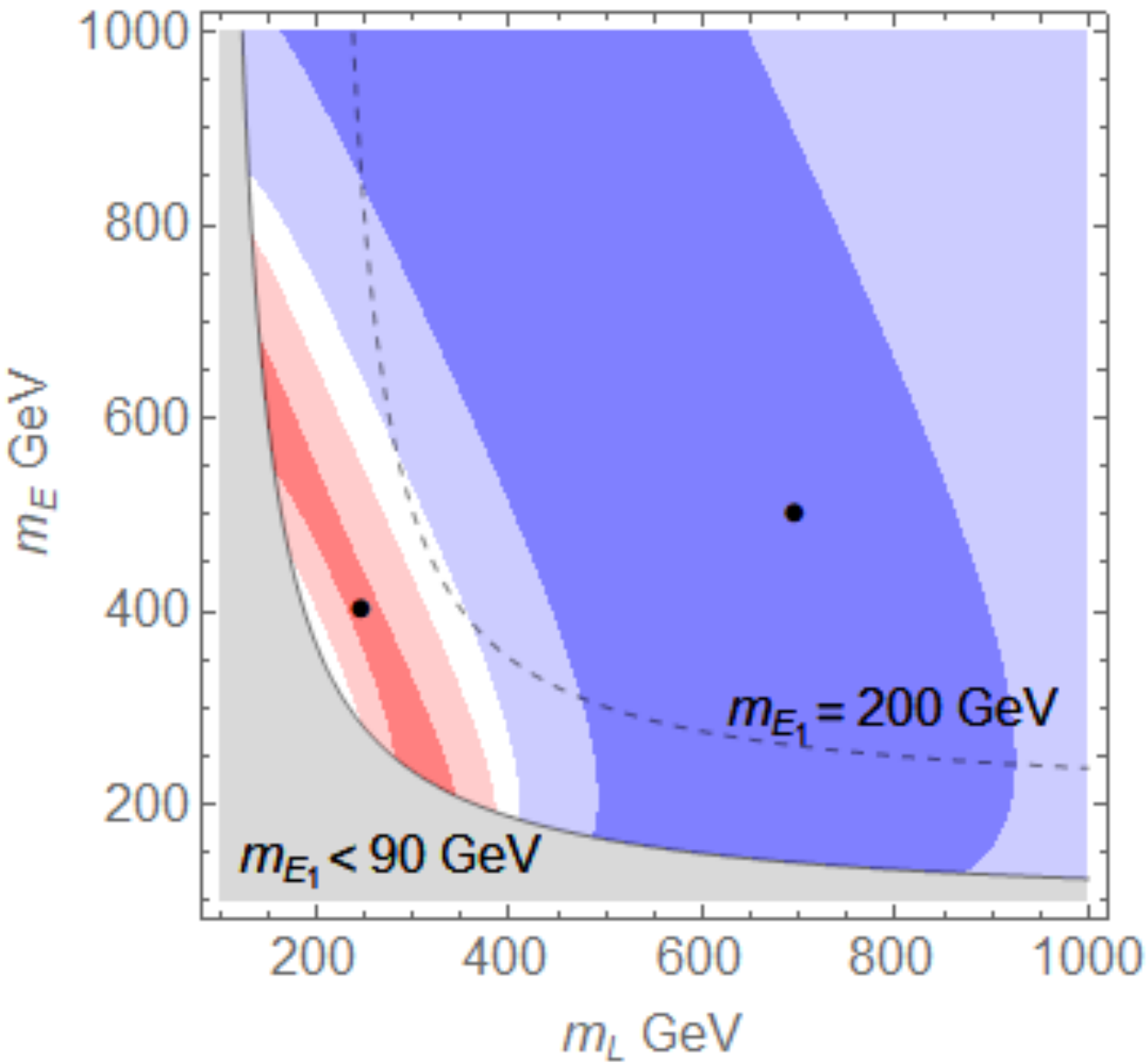} 
\caption{\label{fig-chi2}
Values of $\chi$ from the vector-like leptons. 
}
\end{figure}

\begin{table}[t!]
\centering 
\caption{\label{tab-bench} 
Benchmark points (A) and (B). $\ka=\ol{\ka}=1.0$.  
$\chi_\CDF$ and $\chi_\PDG$ are the values of $\chi$ with the new CDF and PDG results, respectively.  
}
\begin{tabular}{c|cc}\hline
                 & A  & B \\   \hline  \hline 
$(m_L,m_E)~\GeV$ & $(250,400)$ & $(700,500)$ \\   \hline   
$(m_{E_1},m_{E_2})~\GeV$ & $(135.5, 514.5)$ & $(399.3, 800.7)$ \\   
$s_L=s_R$ & $0.549623$ & $0.86553$ \\    
$(S,T,U)$ & $(0.087,0.228,0.012)$ & $(0.019,0.050,0.0004)$ \\ 
$m_W~\GeV$ & 80.4371 & 80.3740 \\ \hline
$(\chi_\CDF, \chi_\PDG)$ & $(0.80,4.00)$ & $(6.03,0.65)$ \\ 
\hline 
\end{tabular}
\end{table}

Figure~\ref{fig-chi2} shows values of $\chi$ from the vector-like leptons, 
$L$ and $E$, for $\ka = \ol{\ka}= 1.0$.  
In the (light) red region, $\chi < 1~(2)$ for the CDF data, 
while the blue region are those for the PDG data. 
We see that the new CDF result favors the light vector-like leptons, 
and $\chi < 2$ is achieved where $m_{E_1}< 200~\GeV$. 
Whereas the PDG result favors heavier vector-like leptons.  
The red region almost coincides with the region 
where the new $m_W$ is explained in Fig.~\ref{fig-ST}. 
The values at the benchmark points are shown in Table~\ref{tab-bench}. 
The point (A) fits to the new CDF result and $\chi= 0.80$, 
while the point (B) fits to the combined PDG result and $\chi= 0.65$. 
These points are plotted as the dots in Fig.~\ref{fig-chi2}.

\subsection{Limits from the collider experiments}

The vector-like lepton should be lighter than 200 GeV 
to explain the new CDF result. 
Such a light vector-like lepton may be excluded by direct searches in the collider experiments. 
If there is no DM particle and a vector-like lepton decays to a muon and a SM boson, 
the Run-1 data at the LHC~\cite{ATLAS-CONF-2013-070} rules out most cases~\cite{Dermisek:2014qca}. 
Furthermore, the recent studies using the Run-2 data
exclude doublet-like vector-like leptons, 
favored to explain the $m_W$ anomaly; 
the lower bound on the mass is about 800 GeV~\cite{CMS:2022nty,CMS:2019lwf}, 
under the assumptions that the lepton flavor is different in these analyses.
Therefore the light vector-like lepton is possibly excluded by the LHC searches. 

In our setup, the vector-like lepton decays to a DM particle and muon. 
In this case, the signal is $2\mu + \met$, where $\met$ comes from the DM.
This signal is studied in the slepton searches at the LHC~\cite{Aad:2019byo,Aad:2019qnd,Aad:2019vnb,Sirunyan:2018nwe,CMS:2020bfa}.
As we have already shown in Ref.~\cite{Kawamura:2020qxo}, 
the limit for the doublet vector-like lepton is about 900 GeV 
if the DM mass is sufficiently light to produce energetic muons. 
The limit is, however, much weaker if $\Delta m := m_{E_1} - m_X \lesssim 100~\GeV$ 
since the muons become soft in this region.   
There are dedicated searches for signals with soft leptons and $\met$ associated with an additional jet~\cite{Aad:2019qnd}. 
If we use the limit on degenerate sleptons~\cite{Aad:2019qnd} as a conservative one, 
the limit is about 250 GeV for $\Delta m \sim 10~\GeV$, 
while the limit is less than 100 GeV for $\Delta m \lesssim 1$ or $\gtrsim 30~\GeV$. 

The muons are too soft to be detected in the detector for even smaller $\Delta m$, 
and hence there is no detectable signal from the vector-like lepton decays. 
In this case, we refer to the limits from the higgsino searches 
since the higgsino has the same quantum number as the doublet vector-like lepton. 
At the LHC, the mono-jet searches~\cite{Barducci:2015ffa,CMS:2017zts,ATLAS:2021kxv}
can not constrain the higgsinos due to its large backgrounds~\footnote{
Recently, it has been proposed 
that mono-$W/Z$ signal may cover light higgsinos $\sim 110~\GeV$
with the full Run-2 data~\cite{Carpenter:2021jbd}.}  
Therefore, the current lower limit of the vector-like lepton
may be about 90 GeV at the LEP experiment~\cite{ALEPH:2002gap,ALEPH:2002nwp}.

To sum up, 
the vector-like lepton with $90<m_{E_1}<200~\GeV$
can still be viable 
if the DM is nearly degenerate with the vector-like lepton. 
Limits in the degenerate region are very sensitive to the mass difference. 
For instance, there will be no limits stronger than $90~\GeV$
from the degenerate slepton searches~\cite{Aad:2019qnd} 
for $\Delta m \lesssim 1~\GeV$ or $\gtrsim 30~\GeV$.  
A dedicated study for the degenerate vector-like lepton 
is an interesting subject but beyond the scope of this paper, 
and thus we only take the LEP bound into account 
while bearing in mind that there may be an exclusion limit on certain mass differences.

\section{Dark matter physics and $\Delta a_\mu$}  
\label{sec-DMM}

The vector-like leptons can participate in the DM thermal production 
via the Yukawa couplings to the SM leptons and DM. 
With the couplings to muon, it also contributes to $\dam$.
We have figured out in Ref.~\cite{Kawamura:2020qxo} that $\dam$ is explained consistently with the DM relic abundance 
if the double and singlet vector-like leptons have a mass mixing. 
In this section, we examine the compatibility of the DM production with the muon $g-2$ and $W$ boson mass anomalies on the two benchmark points in Table \ref{tab-bench}. 

In the thermal freeze-out scenario, 
DM relic abundance is controlled by DM pair annihilation cross section at the freeze-out temperature $T_f \simeq m_X/20$, 
where DM particles can be considered as non-relativistic. 
In the presence of the doublet-singlet mixing, the DM couples to both left and right-handed muons and 
the pair annihilation $XX\to \mu\bar\mu$ has the $s$-wave part, 
\begin{equation}
(\sigma v)_{XX\to \mu\bar\mu} \simeq \frac{(\la_L\la_R)^2}{\pi} \left( \frac{c_R s_L m_{E_1}}{ m_X^2+m_{E_1}^2 } - \frac{c_L s_R m_{E_2}}{m_X^2+m_{E_2}^2} \right)^2 + {\cal O}(v^2),
\end{equation}
where we neglect the muon mass and assume that $\la_L$ and $\la_R$ are real.
Here, $v$ is relative velocity of DM particles. 
This process is expressed in terms of a coupling combination $\la_L \la_R$ like the chirality enhanced contribution in $\Delta a_\mu$.
This suggests that the DM abundance can be highly correlated to $\Delta a_\mu$ in this model.  
In fact, as we discuss in Ref.~\cite{Kawamura:2020qxo} in detail, 
a large new physics contribution $\Delta a_\mu \sim 10^{-8}$ is predicted in this model 
when the $s$-wave part is mainly responsible for the DM production. 
Thus, one needs to make the $s$-wave contribution sub-dominant in the DM annihilation 
by e.g. employing sizable co-annihilation
or introducing a large hierarchy between two coupling constants $\la_{L,R}$ 
to invoke the velocity suppressed $d$-wave annihilation instead of the $s$-wave.

\begin{figure}[t!]
\centering 
\includegraphics[width=0.49\textwidth]{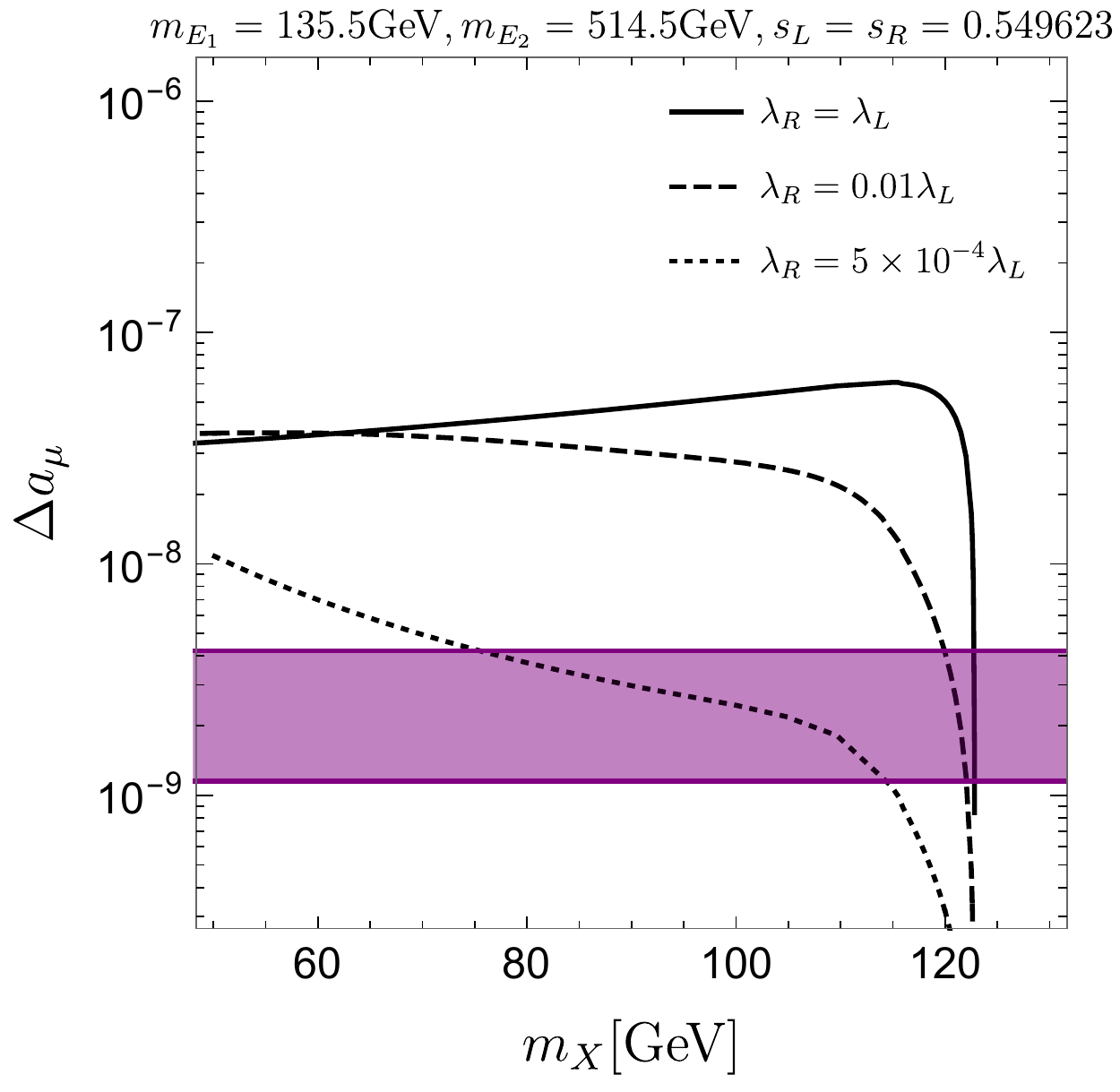} 
\includegraphics[width=0.49\textwidth]{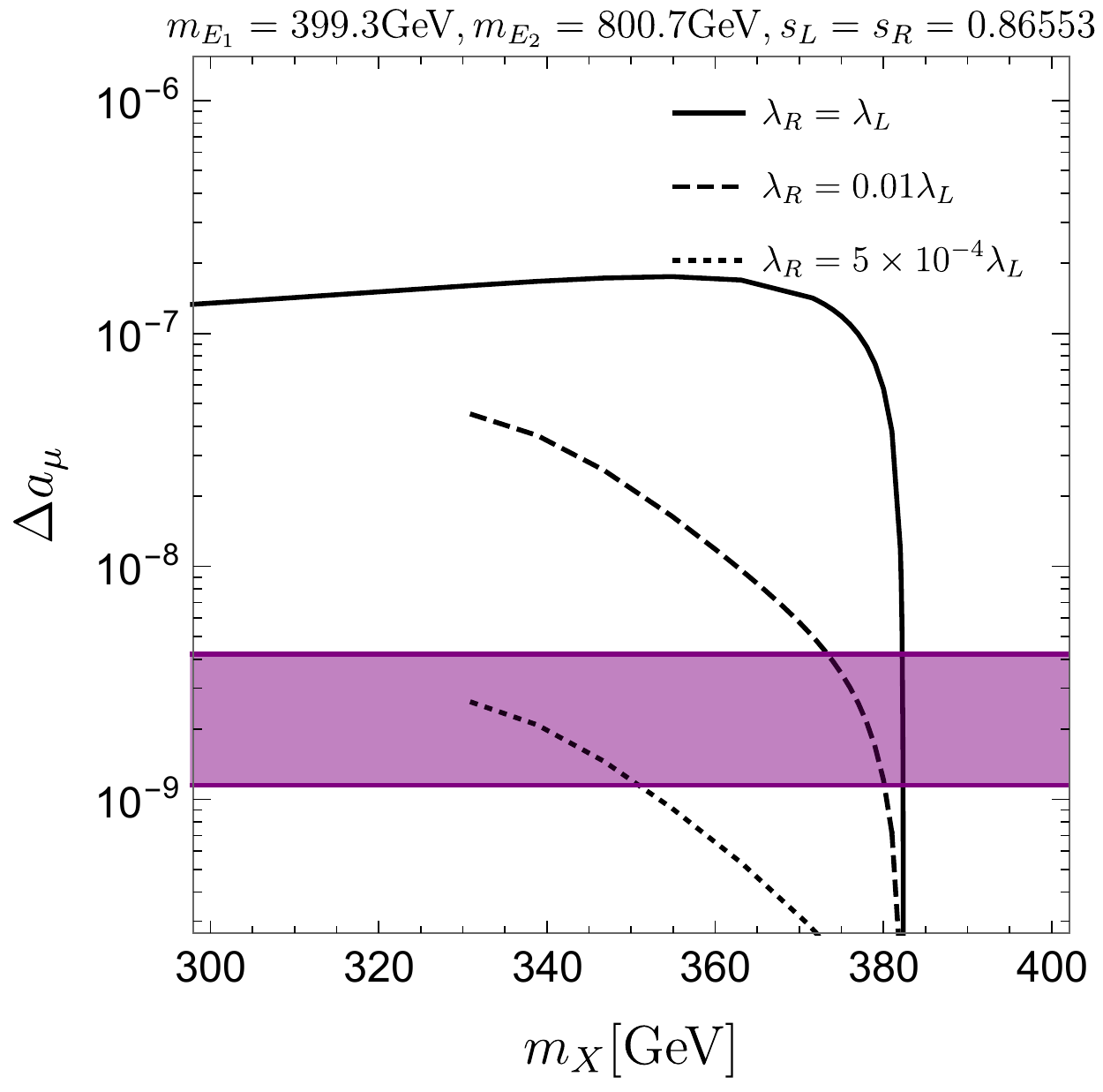} 
\caption{\label{fig-amu}
The predicted value of $\Delta a_\mu$ on the benchmark point A (left) and B (right). 
The purple band represents the 2$\sigma$ favored value of the muon $g-2$. 
For further details, see the text. 
}
\end{figure}

Figure \ref{fig-amu} shows the predicted value of $\Delta a_\mu$ on the benchmark A (left) and B (right).
The black solid, dashed and dotted lines correspond to $\la_R = \la_L, \, 0.01 \la_L$ and $5\times 10^{-4} \la_L$ respectively, 
and the value of $\la_L$ is fixed to explain the observed DM abundance in each case. 
The purple band represents the 2$\sigma$ range of $\dam$.  
We employ {\tt micromegas\_5\_2\_4}~\cite{Belanger:2018ccd} 
to calculate the DM relic abundance including all co-annihilation processes.

We see in Fig.~\ref{fig-amu} that $\dam$ and DM can be consistently explained on both benchmarks 
only if the DM is degenerate to the lighter vector-like lepton $E_1$. 
In this mass region, the DM pair annihilation is sub-dominant and 
the dominant process is the co-annihilation of $E_1$; $E_1\bar E_1 \to \gamma h, WW, ZZ, f\bar{f}$, 
where $f$ is the SM fermions. 
Although all of the co-annihilation processes are comparable, 
$E_1\bar E_1 \to \gamma h$ gives the largest contribution for our benchmark points 
due to the large $\ka$  and $\bar{\ka}$ which are favored to deviate the $T$ parameter. 
Apart from the co-annihilation region, a large $\Delta a_\mu$ is predicted 
due in part to a large Yukawa coupling constant $\la_L$ being required for the DM production, 
and as a result such mass region is strongly disfavored.
It also follows from Fig.~\ref{fig-amu} (left) that 
for $\la_R/\la_L \gtrsim 0.01$, $\Delta m \simeq 10~\GeV$ is favored by $\dam$ and DM,
and the LHC result is likely to exclude the 135 GeV vector-like lepton. 
While, we find $\Delta m \gtrsim 30~\GeV$ to explain $\Delta a_\mu$ if $\la_R/\la_L \sim  5\times10^{-4}$.
In this case, the current LHC limit can be evaded. 
It should be noted, however, that $\la_L$ is sizable with a small $\la_R/\la_L$ and $\la_L\simeq 1.8$ is required for the DM production when $m_X=100$\,GeV and $\la_R/\la_L \lesssim  0.01$.
Otherwise, $\la_L = {\cal O}(0.01-0.1)$ is enough to fit the DM abundance.
In the benchmark (B), the value of $\la_L$ tends to be larger than that of the benchmark (A) mainly because $m_{E_1}$ is heavier.
This tendency is pronounced for a small $\la_R/\la_L$ and, in fact, we find $\la_L$ exceeds the perturbative value $\sqrt{4\pi}$ below $m_X\simeq330$\,GeV when $\la_R/\la_L \lesssim  0.01$. 
That is why the black dashed and dotted lines are interrupted in $m_X\lesssim330$\,GeV.

We briefly comment on direct and indirect detection constraints. 
The DM candidate is a SM gauge singlet and couples only to the muon, 
so DM-nucleon scattering arises at the loop level. 
Furthermore, it is known that in the lepton portal models, 
a real scalar DM starts the scattering at two-loop level via di-photon exchanging\cite{Kopp:2009et}. 
The resulting cross section is highly suppressed and thus well below the reach of the direct detection experiments.
As regards the indirect detection, gamma-ray searches 
at Fermi-LAT~\cite{Ackermann:2013uma,Hoof:2018hyn} and H.E.S.S~\cite{HESS:2013rld}, 
and positron flux measurements at AMS~\cite{Bergstrom:2013jra,AMS:2014xys,AMS:2014bun} 
are important.  
In particular, a real scalar DM with the lepton portal couplings predicts a sharp photon flux 
via virtual internal bremsstrahlung and one-loop processes, 
and can be effectively searched by the Fermi/H.E.S.S line gamma searches~\cite{Toma:2013bka,Giacchino:2013bta,Ibarra:2014qma}. 
If there is no doublet-singlet mixing, 
the bound reads $m_{E_1} \lesssim 1.1~(1.2)~m_X$ 
at best for a purely singlet (doublet) vector-like lepton~\cite{Kawamura:2020qxo}.
Thus, the mass region of our interest may be within the reach of the gamma-ray searches. 
One should note, however, that there is no detailed study on the gamma-ray bound 
in the case with both left and right-handed couplings, 
so that this Fermi/H.E.S.S limit cannot be applied directly to this model.
For the AMS positron measurements, the current conservative lower limit on DM mass is 30\,GeV~\cite{Leane:2018kjk}, 
assuming the cross section of $XX\to\mu\bar\mu$ has the canonical size for thermal production, i.e.
$(\sigma v)_{XX\to \mu\bar\mu}\simeq 3\times10^{-26}\,{\rm cm^3/s}$. 
Hence, the AMS measurements do not provide the relevant limit in this case.

\section{Summary}
\label{sec-summary}

In this paper, we studied corrections to the oblique parameters, 
which are strongly correlated with the $W$ boson mass, 
in the extended SM with the real scalar DM and the vector-like leptons. 
The sizable mixing between the singlet and doublet vector-like leptons 
is crucial to deviate the oblique parameters. 
Since the mixing is induced by the Higgs VEV, 
the vector-like leptons should be close to the EW symmetry breaking scale. 
In fact, we found that the lightest vector-like lepton has to be lighter than 
200 GeV to explain the new CDF result, see Fig.~\ref{fig-chi2}. 
The light vector-like lepton may be excluded by the direct searches at the LHC 
if it decays to the SM particles.
The new lepton, however, can evade the limit 
if it decays to a DM particle whose mass is close to the lepton, 
so that the signals are effectively invisible.

As discussed in Ref.~\cite{Kawamura:2020qxo}, 
the Yukawa couplings involving muon, a real scalar DM and vector-like leptons
can resolve
the discrepancy in the muon anomalous magnetic moment, $\dam$. 
The simultaneous explanation of the DM and $\dam$ 
requires the sizable mixing in the vector-like leptons 
and the mass degeneracy of the lighter vector-like lepton and DM.
Interestingly, the former is required to achieve the shift in $W$ boson mass 
and the latter is to avoid the LHC limits. 
The mass region favored by those anomalies would be covered by the future gamma-ray searches 
utilizing the sharp photon flux from the galactic DM annihilation.

\section*{Acknowledgment}
The work of J.K.
is supported in part by
the Institute for Basic Science (IBS-R018-D1)
and the Grant-in-Aid for Scientific Research from the
Ministry of Education, Science, Sports and Culture (MEXT), Japan No.\ 18K13534. 
%---------------------------------------------------------------------------
S.O. acknowledges financial support from the State Agency for Research of the Spanish Ministry of Science and Innovation through the ``Unit of Excellence Mar\'ia de Maeztu 2020-2023'' award to the Institute of Cosmos Sciences (CEX2019-000918-M) and from PID2019-105614GB-C21 and 2017-SGR-929 grants.
%---------------------------------------------------------------------------
The work of Y. O. is supported by Grant-in-Aid for Scientific research from the MEXT, Japan, No. 19K03867.

\appendix 
\section{Loop functions} 
\label{sec-lfun} 

The loop functions for $\Delta a_\mu$ are defined as 
\begin{align}
 F(x) &\ = \frac{2+3x-6x^2+x^3+6x\log x}{6(1-x)^4},\quad 
 G(x) = \frac{3-4x+x^2+2\log x}{(1-x)^3}.   
\end{align}

The loop functions for the oblique parameters are defined as 
\begin{align}
 \theta_+(y_1,y_2) = y_1 + y_2 - \frac{2y_1y_2}{y_1-y_2} \log\frac{y_1}{y_2}, 
\quad 
  \theta_-(y_1,y_2) = 2\sqrt{y_1y_2} \left( \frac{y_1+y_2}{y_1-y_2} \log \frac{y_1}{y_2}-2 \right), 
\end{align}
and 
\begin{align}
 \chi_+(y_1,y_2) =&\  \frac{y_1+y_2}{2}-\frac{(y_1-y_2)^2}{3} + 
                    \left(\frac{(y_1-y_2)^3}{6}-\frac{1}{2}\frac{y_1^2+y_2^2}{y_1-y_2}
         \right) \log \frac{y_1}{y_2}  \\ \notag 
  + \frac{y_1-1}{6}&\ f(y_1,y_1) + \frac{y_2-1}{6} f(y_2,y_2) 
 + \left(\frac{1}{3}-\frac{y_1+y_2}{6}-\frac{(y_1-y_2)^2}{6} \right) f(y_1,y_2), \\
\chi_-(y_1,y_2) =&\ 
 -\sqrt{y_1y_2} \left[ 
 2+\left(y_1-y_2-\frac{y_1+y_2}{y_1-y_2} \right)\log\frac{y_1}{y_2}  \right. \\ \notag 
 &\ \hspace{2cm} \left. + \frac{f(y_1,y_1)+f(y_2,y_2)}{2} - f(y_1,y_2) 
\right], \\  
 \psi_+(y_1, y_2) =&\ 
 \frac{2y_1+10y_2}{3} + \frac{1}{3} \log\frac{y_1}{y_2}+ 
  \frac{y_1-1}{6} f(y_1,y_1) +   \frac{5y_2 + 1}{6} f(y_2,y_2), \\ 
 \psi_-(y_1,y_2) =&\ 
  - \sqrt{y_1 y_2}\left(4+\frac{f(y_1,y_1)+f(y_2,y_2)}{2} \right).   
\end{align}
Here,  
\begin{align}
\label{eq-deff}
 f(y_1,y_2) = 
\begin{cases}
 \sqrt{d} \log\abs{\dfrac{y_1+y_2-1+\sqrt{d}}{y_1+y_2-1-\sqrt{d}}} & d >  0 \\ 
 0 & d=0 \\ 
 -2 \sqrt{\abs{d}} \left[\tan^{-1} \dfrac{y_1-y_2+1}{\sqrt{\abs{d}}}-
 \tan^{-1} \dfrac{y_1-y_2-1}{\sqrt{\abs{d}}}\right] & d < 0   
\end{cases}
\end{align}
with $d:= (1+y_1-y_2)^2 -4y_1$.  
Note that $\psi_\pm$ are different from those for vector-like quarks shown in Ref.~\cite{Lavoura:1992np}.

{\small
\bibliographystyle{JHEP}
\bibliography{ref_fermionportal}
}

\end{document}